# Pilot Study to Discover Candidate Biomarkers for Autism based on Perception and Production of Facial Expressions

Megan A. Witherow, Norou Diawara, Janice Keener, John W. Harrington, and Khan M. Iftekharuddin

**Purpose:** Facial expression production and perception in autism spectrum disorder (ASD) suggest potential presence of behavioral biomarkers that may stratify individuals on the spectrum into prognostic or treatment subgroups. Construct validity and group discriminability have been recommended as criteria for identification of candidate stratification biomarkers.

**Methods:** In an online pilot study of 11 children and young adults diagnosed with ASD and 11 age- and gender-matched neurotypical (NT) individuals, participants recognize and mimic static and dynamic facial expressions of 3D avatars. Webcam-based eye-tracking (ET) and facial video tracking (VT), including activation and asymmetry of action units (AUs) from the Facial Action Coding System (FACS) are collected. We assess validity of constructs for each dependent variable (DV) based on the expected response in the NT group. Then, the Boruta statistical method identifies DVs that are significant to group discriminability (ASD or NT).

**Results:** We identify one candidate ET biomarker (percentage gaze duration to the face while mimicking static 'disgust' expression) and 14 additional DVs of interest for future study, including 4 ET DVs, 5 DVs related to VT AU activation, and 4 DVs related to AU asymmetry in VT. Based on a power analysis, we provide sample size recommendations for future studies.

**Conclusion:** This pilot study provides a framework for ASD stratification biomarker discovery based on perception and production of facial expressions.

Megan A. Witherow and Khan M. Iftekharuddin, Vision Lab, Dept. of Electrical & Computer Engineering, Old Dominion University; Norou Diawara, Dept. of Mathematics & Statistics, Old Dominion University; Janice Keener, Dept. of Pediatrics, Eastern Virginia Medical School and Developmental Pediatrics, Children's Hospital of The King's Daughters; John W. Harrington, Dept. of Pediatrics, Eastern Virginia Medical School and Children's Medical Group, Children's Hospital of The King's Daughters.

This research study was funded in part by the National Science Foundation under Grant Nos. 1753793, 2139907, and 1828593.

This research study was reviewed and approved by the Institutional Review Boards of Old Dominion University (1424272-21) and Eastern Virginia Medical School (19-06-EX-0152). Written informed consent was obtained from all adult participants and from legal guardians of child participants. Written assent was obtained from all child participants.

Correspondence concerning this paper should be addressed to Khan M. Iftekharuddin, Vision Lab, Dept. of Electrical & Computer Engineering, Old Dominion University, Norfolk, VA, USA. Email: kiftekha@odu.edu

## Introduction

Autism Spectrum Disorder (ASD) is a heterogeneous neurodevelopmental condition. Individuals on the autism spectrum may have a wide variety of behavioral symptom profiles and experiences. In particular, high heterogeneity in the perception and production of facial expressions among autistic individuals compared to neurotypical (NT) individuals has historically resulted in mixed findings regarding the nature of facial expressions in ASD (Keating & Cook, 2020). Rather than diagnostic biomarkers, a recent emphasis on identifying stratification biomarkers to help explain the heterogeneity of ASD has emerged with aims to improve patient care and individualize treatment strategies (Happé & Frith, 2020). Recently, widespread access to the Internet and the ease of technology has enabled ecologically valid measurements of ASD-related behaviors at home, school, and in the community to complement ongoing clinical care, increase access to services, and reduce barriers to research participation (Egger et al., 2018). Behavioral video tracking (VT) and eye tracking (ET) enable low-cost, unobtrusive acquisition of quantitative behavioral measurements of the face and eyes in naturalistic settings (McPartland et al., 2020; Webb et al., 2020), which may be subsequently studied using robust statistical tools for biomarker identification (Acharjee et al., 2020; Hamidi et al., 2023; Shic et al., 2023).

**Related Work**

Historically, ASD research has focused on identifying diagnostic biomarkers that partition individuals into groups of NT and autistic individuals (Keating & Cook, 2020). The role of ASD in perception and production of facial expressions has been studied for decades with conflicting findings, in part due to high heterogeneity in the responses of individuals diagnosed with ASD. The majority of studies report that autistic individuals have greater difficulty (e.g., lower accuracy, higher electroencephalography (EEG) N170 latency) compared to NT controls when completing facial expression recognition tasks, while others report no group differences (Keating & Cook, 2020). Production of facial expressions has been studied via one of two modes of elicitation: (1) spontaneity, i.e., natural elicitation of expressions, or (2) mimicry, i.e., imitation of expressions. Spontaneous expressions of autistic individuals have been reported to be less frequent, shorter in duration, and lower in quality as rated by NT observers (Keating & Cook, 2020). Intensity of spontaneous expressions has been reported to be the same (Drimalla et al., 2021; Keating & Cook, 2020), lesser (Bangerter et al., 2020; Manfredonia et al., 2019), or greater (more exaggerated) (Bangerter et al., 2020; Faso et al., 2015; Quinde-Zlibut et al., 2022; Volker et al., 2009) in individuals diagnosed with ASD compared to NT controls. Some studies (Guha et al., 2015; Samad et al., 2015, 2016) have also reported greater asymmetry in the spontaneous facial expressions of individuals diagnosed with ASD. Regarding facial expression mimicry ability, some studies find no group differences, while others report that expressions posed by autistic individuals are less congruent or less accurate than NT controls (Keating & Cook, 2020).

Recently, focus has shifted from discovery of diagnostic biomarkers to identification of stratification biomarkers that may identify subgroups that are more internally homogeneous (Happé & Frith, 2020). Such stratification biomarkers may identify prognostic subgroups with different tracks of longitudinal symptom development or treatment subgroups of individuals with clinically relevant difficulties, e.g., in social skills, for selective enrollment in interventions (Loth & Evans, 2019). Globally, several large-scale efforts aimed at ASD biomarker discovery have emerged, including European Autism Interventions—A Multicentre Study for Developing New Medications (EU-AIMS) Longitudinal European Autism Project (LEAP) (Loth et al., 2017), Janssen Autism Knowledge Engine (JAKE) (Ness et al., 2016), and the Autism Biomarkers Consortium for Clinical Trials (ABC-CT) (McPartland et al., 2020). In the United States, ABC-CT has resulted in the first two biomarkers, an EEG N170 measure and an ET measure, for ASD to be accepted for evaluation in the United States Food & Drug Administration (FDA) biomarker qualification program (U.S. Food & Drug Administration, 2023).

The highly heterogeneous responses seen in prior studies of facial expressions in ASD suggest that facial expressions may be a meaningful target for stratification. In a study of facial expression recognition with autistic and NT participants ages 14-55 years, Loth et al. (2018) report that 63% of their sample of individuals diagnosed with ASD performed more than two standard deviations below the NT mean, while a smaller subgroup of about 15% of autistic participants performed as well as the NT group. As a part of the EU-AIMS LEAP, Meyer-Lindenberg et al. (2022) investigate facial expression recognition as a candidate stratification biomarker for ASD. In their sample (participants ages 6-30 years and NT or diagnosed with ASD, and/or mild intellectual disability), partitioning individuals diagnosed with ASD into low and high performing subgroups accounts for 9-14% of the variance in ASD-related traits, adaptive behavior, and severity of social difficulties. Furthermore, Meyer-Lindenberg et al. (2022) report neurofunctional differences, i.e., significantly lower functional magnetic resonance imaging (fMRI) activation in the amygdala and fusiform gyrus for the low performing subgroup, between subgroups. Differences in ET to social images, including facial expressions, as a characteristic of ASD has been well documented (Cuve et al., 2018). As a part of ABC-CT, Shic et al. (2022) present the Oculomotor Index of Gaze to Human Faces (OMI), an aggregated measure of ET to faces in videos of social scenes, as a candidate stratification biomarker for ASD. OMI has recently been accepted into the FDA's biomarker qualification program (U.S. Food & Drug Administration, 2023). While OMI focuses on social scenes overall and not on facial expressions in particular, recent ET studies on facial expression recognition have also found reduced visual attention to the face by participants diagnosed with ASD compared to NT controls, as well as greater difficulty in recognizing negative or complex emotions such as ‘anger’, ‘disgust’, and ‘fear’ (Su et al., 2018; Tsang, 2018).

For measuring facial expression production, the Facial Action Coding System (FACS) (Ekman et al., 2002) is considered the gold standard. FACS assigns action unit (AU) codes to the individual muscular movements of the human face (Ekman et al., 2002). Quinde-Zlibut et al. (2022) use FACS AUs to measure spontaneous facial expression production among adult (ages 18-59 years) autistic and NT participants for subgroup identification. For the ASD group, they report that increased expressiveness is associated with poorer facial expression recognition performance (Quinde-Zlibut et al., 2022). Moreover, Quinde-Zlibut et al. (2022) have identified a subgroup of autistic adults in their sample that show heightened expressiveness (‘engagement’ summary score from the commercially available iMotions AFFDEX software (https://imotions.com/products/imotions-lab/modules/fea-facial-expression-analysis/), computed as the average activation of upper and lower face AUs). Bangerter et al. (2020) study the spontaneous facial responses of autistic and NT individuals ages 6 to 63 in response to funny videos. Using the iMotions FACET software (no longer commercially available), two constituent AUs of a smile / ‘happy’ expression are measured: AU 6 and AU 12 (Bangerter et al., 2020). They report that, on average, individuals diagnosed with ASD show less activation of AUs 6 and 12 as compared to the NT control group (Bangerter et al., 2020). However, within the ASD group, they identify “over-responsive” and “under-responsive” subgroups, which display more intense and less intense responses to the stimuli, respectively, as compared to the NT group (Bangerter et al., 2020). Using JAKE, Manfredonia et al. (2019) study facial expression mimicry in individuals ages 6-54 years, including an ASD group and an NT group. They study activation of two AUs, as measured using the iMotions FACET software, per each of six facial expressions: ‘anger’ (AUs 4 and 23), ‘disgust’ (AUs 9 and 10), ‘fear’ (AUs 4 and 5), ‘happy’ (AUs 12 and 20), ‘sad’ (AUs 1 and 15), and ‘surprise’ (AUs 5 and 26) (Manfredonia et al., 2019). Manfredonia et al. (2019) report statistically significant differences between autistic and NT individuals’ portrayals of ‘happy’ (AU 12), ‘fear’ (AU 5), ‘surprise’ (AU 5), and ‘disgust’ (AU 9). They also find significant negative correlations between some AUs (‘happy’ AU 12 and ‘fear’ AU 5) and social communication scores (Manfredonia et al., 2019). Although the purpose of their study is not to identify subgroups, Manfredonia et al. (2019) report that in some cases, more activation of AUs corresponded to a greater severity of symptoms, which may suggest a subgroup of individuals with more exaggerated or intense expressions. Drimalla et al. (2021) investigate recognition and mimicry of facial expressions in a sample of autistic and NT individuals ages 18 to 62 years. They report significantly higher recognition accuracy and faster response times for the NT group compared to the ASD group (Drimalla et al., 2021). To automatically recognize AUs during expression mimicry of AU-labeled images of actors posing expressions, Drimalla et al. (2021) use the open-source OpenFace 2.0 (Baltrusaitis et al., 2018) software. Drimalla et al. (2021) report that the imitated expressions of participants diagnosed with ASD are significantly more different than the stimulus expressions, with significantly more variance in intensity, and require significantly more time to pose when compared to the NT control group. Furthermore, more accurate imitation of facial expressions is found to be positively associated with better recognition performance (Drimalla et al., 2021).

In addition to individual heterogeneity across the spectrum, task design and participant characteristics play an important role in the elicitation and measurement of ASD-related behaviors. Facial expression stimuli may be presented as either static, i.e., still images, or dynamic, i.e., videos or animations. Both static and dynamic expressions have been shown to elicit differential responses, e.g., reduced recognition accuracy, in individuals diagnosed with ASD as compared to NT controls (Keating & Cook, 2020). Keating and Cook (2020) point out a need for both static and dynamic expression stimuli to be used in studies of ASD, as it is unclear to what extent autistic individuals may rely on static features (e.g., configuration of the face) versus dynamic features (e.g., order and speed of moving facial muscles) during processing of facial expressions. Effective task design also requires consideration of participant engagement. It has been noted that many individuals on the spectrum have an affinity and interest in technology, including increased engagement with 3D avatar characters (Kellems et al., 2023; Putnam et al., 2019). In an ET study on the visual processing of real and avatar faces by children diagnosed with ASD, Pino et al. (2021) have found that participants show increased interest and more visual exploration of the avatar faces compared to the real faces. Furthermore, customization of avatars has been recommended as an important design consideration for the development of interactive technologies for individuals on the spectrum (Bozgeyikli et al., 2018; Putnam et al., 2019) and has been shown to increase task engagement and enjoyment among both NT and autistic individuals (Lee et al., 2023). Individual participant characteristics such as age, gender, and intelligence quotient (IQ) may also affect behavioral responses (Keating & Cook, 2020). Alexithymia, a subclinical personality trait characterized by difficulty in describing one's own emotions and with a prevalence of approximately 50% of the autistic population (Kinnaird et al., 2019) and 10% of the NT population (Goerlich, 2018), has also been found to affect perception and production of facial expressions in both NT and autistic individuals (Keating & Cook, 2020).

Large scale studies aimed at biomarker qualification, such as ABC-CT in the United States and EU-AIMS LEAP in Europe, have established a thorough framework for evaluation of potential biomarkers. To identify biomarkers, ABC-CT recruits hundreds of participants for multi-day studies including diagnostic confirmation, data acquisition, and deep phenotyping (McPartland et al., 2020; Webb et al., 2020). Numerous candidate measurements, or dependent variables (DVs), are evaluated using biomarker qualification criteria based on assay validity, data acquisition rates, distributional properties in the NT group, and replication in an independent sample (Webb et al., 2020). The DVs included in studies like ABC-CT have been discovered and previously published in smaller scale studies (Webb et al., 2020). These DVs are selected for inclusion as candidate biomarkers based on assay validity, which includes two criteria: construct validity and group discriminability (Webb et al., 2020). Construct validity ensures that the experimental task elicits the intended response in the NT control group, and group discriminability confirms the presence of differences between autistic and NT participant groups (Webb et al., 2020). According to Shic et al. (2022), group discriminability in the context of stratification is not expected to have effect sizes with diagnostic precision but rather, indicate broad group-level (ASD or NT) distributional differences associated with more homogeneous subgroups within the ASD group.

Online research studies may play a major role in ASD research by providing online access to large numbers of participants and longitudinal follow up (Feliciano et al., 2018). For example, the Simons Foundation Powering Autism Research for Knowledge (SPARK) (Feliciano et al., 2018) now has over 100,000 people diagnosed with ASD and 175,000 of their family members (Simons Foundation, 2024) sharing medical and behavioral information online through questionnaires and mailing in saliva for genetic analysis. The rise of online research has motivated platforms for remote data collection, such as Apple ResearchKit (apple.com/researchkit) which has been applied to ASD research for collection of behavioral data in the Autism & Beyond research study (Egger et al., 2018). The design of pilot studies compatible with the online model is an important step towards scalability for future large-scale replication and validation of research findings.

There may be potential challenges associated with online research. Recently, it has been reported that online qualitative research studies (e.g., focus groups and interviews) have seen a rise in 'scammer participants' attempting to pose as autistic individuals or their parents (Pellicano et al., 2024). Some characteristics of suspect participants include appointment booking data suggesting that they are in different countries than they claim to be in, keeping cameras off during Zoom/Teams interviews, brief and vague responses, discrepancies in responses (e.g., names and ages changing), frequent inquiries about payment, etc. (Pellicano et al., 2024). However, numerous strategies, including careful screening over telephone or videoconferencing, requiring the webcam to be turned on at the beginning of the research session, and checks to ensure that participants are within the geographic limits of the study (e.g., checking the time zone of the appointment booking) may help safeguard data integrity of such studies (Pellicano et al., 2024). Another challenge of online research is diagnostic confirmation. For example, SPARK requires its autistic participants to have received a lifetime professional diagnosis of ASD. Although diagnoses in the SPARK cohort are not independently verified, study of the validity of self- and caregiver-reported diagnoses has shown good agreement with diagnosis of ASD based on electronic medical records (Fombonne et al., 2022). While telehealth research reports that autistic individuals tend to be more comfortable and relaxed interacting with clinicians from their homes, there may be a need for more parent involvement when interacting with young children (Gibbs et al., 2021). It has also been noted that during online videoconferencing, some autistic individuals may find their own webcam video distracting (Gibbs et al., 2021). Screening participants for suitability and informing participants/parents of what to expect beforehand has been recommended as some children with moderate or severe challenging behaviors may be less likely to stay engaged during online interactions with clinicians (Gibbs et al., 2021).

## Contributions

Motivated by recent progress into the discovery and qualification of stratification biomarkers for ASD, we have conducted an IRB-approved pilot study on facial expression perception and production among children and young adults diagnosed with ASD compared to age- and gender-matched NT controls. Prompted by successful online research studies in the literature (Egger et al., 2018; Feliciano et al., 2018; Simons Foundation, 2024), and the global COVID-19 pandemic this pilot study has been conducted online with steps taken to ensure the integrity of the data and comfort of participants (Gibbs et al., 2021). Participants have completed recognition and mimicry tasks using previously validated static and dynamic stimuli based on customizable 3D avatars (Witherow et al., 2024) while their webcam captures ET and VT of the face. Since the avatar stimuli are labeled with AUs, we are able to define constructs for expression mimicry based on the avatar AUs. It has been shown that facial expression analysis models that are trained using adult expressions (such as OpenFace 2.0 and iMotions) may perform poorly on child facial expressions (Witherow et al., 2023; Witherow et al., 2020). Therefore, we use state-of-the-art deep neural network models (Witherow et al., 2024; Witherow et al., 2023), that have undergone domain adaptation for use in our age group (children and young adults, ages 8 to 20 years) (Ganin & Lempitsky, 2015; Witherow et al., 2023), to extract facial expression and AU labels from the webcam images for behavioral VT. Furthermore, we measure the asymmetry of facial expressions in ASD, which has been investigated by few prior studies (Guha et al., 2015; Samad et al., 2015, 2016). We evaluate our DVs (e.g., participants' facial expressions, activations and asymmetry of AUs, ET measurements, etc. in response to different avatar-rendered facial expressions) using ABC-CT's criteria of construct validity and group discriminability in order to identify candidate stratification biomarkers for future study. Given the large number of parameters designed to capture the AUs, and their complex structures, the use of statistical methods becomes useful to determine the functional forms of the interactions and build group classifications, e.g., ASD vs NT, for group discriminability. The methods we propose are built from the Boruta algorithm models (Acharjee et al., 2020; Kaneko, 2021; Kursa et al., 2010) that circumvent unrealistic assumptions of normality and independence in order to capture and showcase class behaviors. We identify one candidate biomarker plus fourteen additional DVs that may be of interest for future research and provide sample size recommendations for future studies.

# Methods

## Participants

Participants include English-speaking children and young adults between ages 8 to 20 years and residing in the United States. All participants are required to be generally healthy, with either no diagnosis of mood disorders or no change in medication regimen for six months, report being able to sit and interact with a computer for one hour, and have an intelligence quotient (IQ) of 70 or above. All participants complete the Kaufman Brief Intelligence Test Second Edition (KBIT-2) (Kaufman & Kaufman, 2004) for assessment of IQ.

Two groups have been recruited: an NT group and a group of individuals who have received a diagnosis of ASD (henceforth, ASD group). Participants for the ASD group have been recruited through flyers distributed nationwide by affiliates of the Autism Society across the continental United States; the Autism Science Foundation; the Organization for Autism Research; Parents of Autistic Children of Northern Virginia; a hospital; and autism-related Facebook groups. The NT group has been recruited through flyers posted at the research group's website, a university, public libraries, and community centers.

To confirm that all eligibility criteria are met and to safeguard against possible scammer participants, a phone screening interview has been conducted with each participant or their parent/guardian via their United States-based telephone number. Following screening, participants are asked to provide their time zone during scheduling. The time zones are confirmed based on timestamps in scheduling emails. Given the sensitive nature of diagnostic records (Pellicano et al., 2024), participants are invited, but not required, to provide the research team with documentation of diagnosis.

Written informed consent is obtained from all adult participants and from the legal guardians of child participants. All child participants have provided their written assent to participate in the research study. Adult participants and legal guardians of child participants are also invited to complete an alexithymia assessment about the participant, either the Bermond-Vorst Alexithymia Questionnaire (BVAQ) (Vorst & Bermond, 2001) for adults or Children's Alexithymia Measure (CAM) (Way et al., 2010) for children, for which separate informed consent is obtained. Each participant is compensated with a $10.00 Visa gift card (valid only in the United States), emailed to the participant's email address on file.

## Experimental Protocol

Motivated by successful online research studies (Egger et al., 2018; Feliciano et al., 2018; Simons Foundation, 2024), we design an online study protocol to conduct during the COVID-19 pandemic. Experiments have been conducted over a Zoom call between the participant (and participant's parent/guardian if the participant is under 18 years old) and research team. The call begins with the webcam turned on in Zoom. A researcher checks that the participant is centered in front of their webcam and instructs the participant or parent/guardian to navigate to a web URL where the experimental tasks are hosted. Once at the web URL, the webcam is disconnected from Zoom and the participant or parent/guardian is instructed to give the website permission to access the webcam feed. To prevent participants from being distracted by their own webcam video (Gibbs et al., 2021), no visual webcam feed is shown during the experimental tasks. Participants complete recognition (REC) and mimicry (MIM) tasks using the validated, customizable avatars as previously described by Witherow et al. (2024). During the REC task, the participant is asked to click the button for which of six expressions ('anger', 'disgust', 'fear', 'happy', 'sad', or 'surprise') they recognize as being shown on the avatar's face. During the MIM task, the participant is asked to pose the same expression as the avatar in front of their webcam. Given possible variations in engagement and responses to static or dynamic stimuli, each task is completed under four conditions: uncustomized avatar with static expressions (US), uncustomized avatar with dynamic expressions (UD), customized avatar with static expressions (CS), and customized avatar with dynamic expressions (CD). The order of tasks and conditions is REC-US, MIM-US, REC-UD, and MIM-UD, followed by an avatar customization screen, and then REC-CS, MIM-CS, REC-CD, and MIM-CD. Conditions involving the uncustomized avatar are completed first to avoid biasing participants' responses to the uncustomized avatar (e.g., due to disappointment) after having created their customized avatar. For each task and each condition, each of the six expressions ('anger', 'disgust', 'fear', 'happy', 'sad', 'surprise') is shown twice for a total of 12 trials. The order of expressions is randomized within each task and condition. WebGazer.js (https://webgazer.cs.brown.edu/) (Papoutsaki et al., 2016) is used to record video frames for facial VT and webcam-based ET fixation coordinates from the participants' webcams. Participants are informed that they may take a break at any time.

## Data Acquisition Rates

To study acquisition rates for different types of data (e.g., VT, ET) and patterns of possible data loss (e.g., due to participants moving out of frame, ET track loss, etc.), we report the percentage of missing values for each group per task, condition, and expression.

## Derivation of DVs

Our DVs may be described using (stimulus, measurement) pairs. Stimulus refers to a particular expression ('anger', 'disgust', 'fear', 'happy', 'sad', 'surprise') under a particular stimulus condition (US, UD, CS, or CD) presented during a particular task (REC or MIM). The measurements are defined based on ET, VT, and button click data collected from the participants. There are 312 DVs in total as follows. During the REC task, two types of measurements are collected to include:

a) The participants' recognition accuracy (%Acc) is calculated based on the percentage of correct responses (clicking the button labeled with the name of the expression that is shown on the avatar).
b) Following Shic et al. (2022), %Gaze Face is computed as the percentage of participants' ET fixation duration gazing at the avatar's face.

The total number of DVs for the REC task is *2 for %Acc and %Gaze Face × 6 expressions × 4 stimulus conditions = 48*. During the MIM task, four different types of measurements are collected as follows:

a) The age-appropriate facial expression classification model described by Witherow et al. (2023) is used to predict the participants' mimicked expressions from VT. The model outputs softmax probabilities (range 0 to 1) for each expression ('anger', 'disgust', etc.). The softmax probability corresponding to the stimulus expression is used as a measurement of the participants' ability to mimic the avatar's overall expression (EXPR measurement).
b) The participants' ability to mimic each AU presented by the avatar is quantified by the predicted AU activation (ACT measurement, range 0 to 1) in VT frames using an AU model adapted from Witherow et al. (2024).
c) The left-right asymmetry of the participants' AU activations (ASYM measurement) is computed as the difference of left-right activations for the AU in VT frames as predicted by the AU model adapted from Witherow et al. (2024).

d) The %Gaze Face is computed based on ET in the same way as for the REC task.

The total number of DVs for the MIM task is *(2 for EXPR and % Gaze Face × 6 expressions + 2 for ACT and ASYM × (7 AUs in 'anger' + 2 AUs in 'disgust' + 7 AUs in 'fear' + 2 AUs in 'happy' + 4 AUs in 'sad' + 5 AUs in 'surprise')) × 4 stimulus conditions = 264. 48 REC DVs + 264 MIM DVs = 312 total DVs.* We use Ganin and Lempitsky (2015)'s unsupervised domain adaptation method to adapt the AU model described by Witherow et al. (2024) for our age group by finetuning the network on facial expression samples collected in this study for 50 epochs with a leaning rate of 1e-7. Using the adapted model, ACT and ASYM values are obtained for 16 AUs (AUs 1, 2, 4, 5, 6, 7, 10, 11, 12, 15, 17, 20, 23, 25, 26, 27). We specify the DVs as (stimulus, measurement) pairs such as (MIM-US 'Happy', VT ACT AU 6) or (REC-CD 'Sad', ET %Gaze Face).

## Constructs

Following ABC-CT (McPartland et al., 2020; Shic et al., 2022; Webb et al., 2020), we evaluate the construct validity of each DV based on whether the expected response is elicited in the NT group and use one sample t-tests to test for validity. The construct for %Acc based DVs during REC is intact expression recognition, with null hypothesis $H_0$: μ=16.7% and alternative $H_a$: μ >16.7%, i.e., the NT mean for %Acc is greater than chance (1 clicked expression button / 6 total expression buttons = 16.7%). The construct for DVs measuring %Gaze Face during REC and MIM is gaze preference to the face, with $H_0$: μ =15.0% and $H_a$: μ >15.0%, i.e., the NT mean for %Gaze Face is greater than random gaze (the face occupies 15.0% of the visual scene) (Shic et al., 2022). Intact expression mimicry is the construct for DVs measuring EXPR and ACT of AUs during the MIM task. For EXPR, $H_0$: μ =16.7% and $H_a$: μ >16.7% (greater than chance). For ACT of AUs, $H_0$: μ =0 and $H_a$: μ >0 (AU is present). The expected response in the NT group for ASYM of AUs is symmetrical AU activation. We consider $H_0$: μ =0 (symmetrical activation) and $H_a$: μ ≠0 (asymmetrical activation). For all DVs except those based on ASYM of AUs, the construct is valid if $H_0$ is rejected. For ASYM of AUs, we consider the construct valid if the corresponding construct for ACT of the AU is valid and we fail to reject $H_0$.

## Imputation of Missing DVs

To impute any missing DVs, we evaluate the performance of five different imputation methods on our data set using samples with no missing values. These five methods include simple imputation with a) the mean or b) the median value of the DV, multiple imputation by chained equations (MICE) (Slade & Naylor, 2020; van Buuren & Groothuis-Oudshoorn, 2011) using c) Bayesian ridge regression or d) random forest regression, and e) k-nearest neighbors (KNN) imputation (Troyanskaya et al., 2001).

We follow prior studies (Slade & Naylor, 2020; Toure et al., 2023) to design an experiment to compare among the five imputation methods. From the full data set, we determine the set of DVs that are missing for one or more samples. Next, we identify a reduced data set of samples with no missing DVs. Then, we repeat the following procedure for each target DV in the set of all missing DVs as follows:

a) We consider leave one out cross validation (LOOCV) of the reduced data set (samples with no missing values) to obtain train/test splits of the data. In LOOCV, each sample serves as the test set once and remaining samples form the train set.
b) For each split, we save the ground truth value of the target DV from the test sample. Then, we assign 'not a number' ('NaN', denotes missing) to the target DV in the test sample.
c) For each sample in the train set, we randomly assign DVs from the set of all missing DVs to 'NaN' with a probability of 50%. This results in a train set with 50% missing values among DVs from the set of missing DVs.
d) We perform imputation with each of the methods.
e) We repeat steps b-d for each of the train/test splits. Using the stored ground truth values and imputed values, we compute evaluation metrics: mean squared error (MSE), root mean squared error (RMSE), and mean absolute error (MAE).

We use Scikit-learn (https://scikit-learn.org/) for our implementation. For MICE, we use mean imputation as the initial strategy and obtain the imputed values by averaging over ten repeated imputations. We use the default settings for Bayesian ridge regression. For random forest, we use ten trees in the ensemble. For KNN, we consider the five nearest neighbors. We average the evaluation metrics over all missing DVs to obtain aggregated metrics for performance comparison among imputation methods. We use the best performing method to impute missing values in the full data set.

## Group Discriminability

Among the DVs with valid constructs, we use the Boruta method (Kursa et al., 2010) to find all DVs that are relevant in discriminating between the NT and ASD groups. For Boruta method, we consider all DVs with valid constructs as features and group (ASD or NT) as the classification labels. Originally developed with genetics research in mind, Boruta is an 'all-relevant' feature selection method that is appropriate for handling correlated features (Kursa et al., 2010), as is expected to be the case with our DVs. For example, ACT AU 6 and ACT AU 12 are expected to occur together during the mimicry of a 'happy' expression (Ekman et al., 2002). Boruta offers numerous advantages: identification of all discriminative features given a specified Type I error rate α, high stability of feature selections, and because it is based on trees, no assumption on the distribution of the data (Acharjee et al., 2020; Kursa et al., 2010). Further, the Boruta method does not require sample independence and normality assumptions.

Boruta is implemented as a wrapper around the random forest classification algorithm, an ensemble method comprising decision trees independently developed on different bootstrapped samples of the data (Kursa et al., 2010). Each tree in the forest assigns an importance value to each feature in the tree based on its contribution to the classification loss (Kursa et al., 2010). To calculate the importance of each feature in the forest, the average loss for the feature among all trees in which it is present may be divided by its standard deviation to generate a Z score (Kursa et al., 2010). These Z scores are used to measure feature importance in the Boruta algorithm (Kursa et al., 2010). Then, Boruta works as follows (Kursa et al., 2010):

a) The information system is extended by making copies of all of the features, called 'shadow features'. To remove correlations with the classification labels, the sample values within each of the shadow features are permuted.
b) The random forest algorithm is run on the extended information system to determine the importance values for the features and shadow features.
c) The percentile (*perc*) of the shadow features' importance is used as a reference value. Features that have a higher importance than the reference value are assigned a 'hit'. In standard Boruta, *perc*=100, so features must have higher importance than the most important shadow feature to be assigned a 'hit'.
d) Steps a-c are repeated for a specified number of iterations or until all features are deemed 'important' or 'unimportant'. After each iteration, p-values are computed using the binomial distribution, e.g., a feature is assigned a 'hit' k times in n iterations (Bernoulli trials) with the null hypothesis that the probability of a 'hit' p is 0.5, i.e., $H_0$: p=0.5. Two one-tailed binomial tests are performed: a test of rejection ($H_a$: p<0.5) and a test of confirmation ($H_a$: p>0.5). In the test of confirmation, we consider features with a Bonferroni-corrected p-value of less than α=0.05 as 'important'. Similarly, features with a Bonferroni-corrected p-value of less than α=0.05 per the test of rejection are considered 'unimportant'.

When the number of samples is small, there is a greater likelihood that the permuted shadow features are correlated with the classification labels by chance (Kaneko, 2021). To address this issue, a modified version of Boruta, called r-Boruta (Kaneko, 2021), that adjusts perc to account for this chance correlation may be used. The new value of *perc* is determined by generating a large number of random features (100,000 in our case), computing the correlation coefficients between the random features and classification labels, and taking the maximum absolute value times 100 as *perc* (Kaneko, 2021). We use the BorutaPy library (https://github.com/scikit-learn-contrib/boruta_py) implementation of Boruta and r-Boruta. Following BorutaPy's documentation, we set the maximum tree depth to 5, allow BorutaPy to dynamically adjust the number of trees, and set the maximum number of iterations to 1000.

## Power Analysis

We use Acharjee et al. (2020)'s PowerTools framework for power analysis of candidate biomarkers. PowerTools uses the observed effect size for each DV to generate multiple synthetic data sets following a series of sample sizes (e.g., 22, 44, 88, …). The associated statistical power for each sample size is reported.

# Results

## Participant Characteristics

Twenty-two participants (11 in each group) are included in our final analysis. ASD and NT groups are selected from a total of 32 volunteers (11 diagnosed with ASD and 21 NT) who completed the study. The ASD group includes all 11 participants diagnosed with ASD. The NT group of 11 participants is formed by matching NT participants with participants in the ASD group on age (±1 year) and gender. Eight out of eleven participants in the ASD group have provided documentation of diagnosis. CAM (Way et al., 2010) and BVAQ (Vorst & Bermond, 2001) alexithymia scores are standardized based on the population values reported in their respective publications. We obtain alexithymia scores for all participants except two in the ASD group, which we impute using KNN imputation. Participant characteristics are summarized in Table 1.

Shapiro-Wilk tests show that the data collected from the participants exhibit possible departure from normality and may have extreme values or may be thought of as a mixture of feature characteristics. Due to time dependence associated with ordering of tasks, the data collected includes correlation and does not reflect complete independence. To further explore these observations about the data, we propose the deep neural network models and Boruta/r-Boruta models that are appropriate for data with correlation and do not assume independence or normality of the data.

### Data Loss

Table 2 reports the percentage of missing samples by task and data modality (REC ET REC button clicks, MIM ET, and MIM VT). One participant in the ASD group (9.09%) and two participants in the NT group (18.18%) do not have any ET samples due to technological issues (e.g., incompatible hardware, low Internet bandwidth). Additional ET samples are lost due to track loss, e.g., due to participants moving out of range. The primary reason for loss of VT is participants leaning in too close to the camera (cutting off their lower face). The greatest data loss is seen for the CD condition, which occurs at the end of the experimental session, likely due to participants moving out of their original calibrated position. To address data loss, we impute missing values using the robust KNN imputation approach, which provides the best performance in the imputation study with much higher levels of simulated data loss (50%) than observed in the experimental data.

### Construct Validity

A total of 220 out of 312 DVs have a valid construct. Table 3 summarizes the results for the tests of construct validity. A construct may be invalid either because the task does not elicit the expected response in the NT group and/or we are unable to measure the elicited response. We further validate these findings citing relevant literature in the discussion section.

### Imputation Study

Comparison of five imputation methods (mean imputation, median imputation, MICE using Bayesian ridge regression, MICE using random forest, and KNN imputation) is carried out with nine participants that have no missing data. Table 4 reports mean MSE, RMSE, and MAE metrics, averaged over all DVs in the missing set. The missing set consists of DVs that have one or more missing samples in the full data set, e.g., 20 or 21 samples instead of the full 22 samples. There are 158 DVs in the missing set. We note that 158 DVs may appear more inflated than actuality as loss of VT samples, e.g., due to a participant leaning in too close to the webcam, may affect multiple DVs. For example, VT for MIM of ‘anger’ during any stimulus condition (US, UD, CS, CD) is associated with 7 ACT DVs, 7 ASYM DVs, and 1 EXPR DV. Therefore, loss of one VT frame during MIM of ‘anger’, e.g., under the US

Table 1
*Participant Characteristics*

| Characteristic | ASD Group | NT Group | t-statistic | p-value |
|---|---|---|---|---|
| Number of participants | 11 | 11 | -- | -- |
| Gender (N Males, N Females) | 8, 3 | 8, 3 | -- | -- |
| %Male | 72.7% | 72.7% | -- | -- |
| Age in years (M, SD) | 14.09 (4.44) | 14.00 (4.05) | 0.0502 | 0.9605 |
| KBIT Full-scale IQ (M, SD) | 100.09 (15.16) | 115.45 (12.48) | -2.5950 | 0.0173* |
| KBIT Verbal IQ (M, SD) | 98.27 (14.42) | 110.54 (9.08) | -2.3882 | 0.0269* |
| KBIT Nonverbal IQ (M, SD) | 101.54 (20.04) | 115.72 (17.68) | -1.7600 | 0.0937 |
| Standardized Alexithymia Score (M, SD) | 0.0103 (0.5315) | -0.3427 (0.9709) | 1.3385 | 0.1995 |
| *Imputed (M, SD)* | *-0.0057 (0.4829)* | | *1.0308* | *0.3150* |

*Note.* ‘*’ indicates statistical significance with $\alpha = 0.05$. M stands for the mean and SD for the standard deviation. KBIT stands for Kaufman Brief Intelligence Test, Second Edition.

Table 2

*Percentage of Missing Data by Task, Condition, and Expression for ASD and NT Groups*

| Expression | Condition | REC ET ASD | REC ET NT | REC Button Clicks ASD | REC Button Clicks NT | MIM ET ASD | MIM ET NT | MIM VT ASD | MIM VT NT |
|---|---|---|---|---|---|---|---|---|---|
| surprise | CD | 18.18% | 27.27% | 0.00% | 0.00% | 36.36% | 36.36% | 27.27% | 9.09% |
| | CS | 9.09% | 18.18% | 0.00% | 0.00% | 9.09% | 18.18% | 9.09% | 0.00% |
| | UD | 9.09% | 18.18% | 0.00% | 0.00% | 9.09% | 18.18% | 18.18% | 0.00% |
| | US | 9.09% | 18.18% | 0.00% | 0.00% | 9.09% | 18.18% | 0.00% | 0.00% |
| sad | CD | 18.18% | 27.27% | 0.00% | 0.00% | 36.36% | 36.36% | 27.27% | 0.00% |
| | CS | 9.09% | 18.18% | 0.00% | 0.00% | 9.09% | 18.18% | 9.09% | 0.00% |
| | UD | 9.09% | 18.18% | 0.00% | 0.00% | 9.09% | 18.18% | 9.09% | 0.00% |
| | US | 9.09% | 18.18% | 0.00% | 0.00% | 9.09% | 18.18% | 0.00% | 0.00% |
| happy | CD | 18.18% | 36.36% | 0.00% | 0.00% | 36.36% | 27.27% | 27.27% | 0.00% |
| | CS | 9.09% | 18.18% | 0.00% | 0.00% | 18.18% | 27.27% | 9.09% | 0.00% |
| | UD | 9.09% | 18.18% | 0.00% | 0.00% | 9.09% | 18.18% | 18.18% | 0.00% |
| | US | 9.09% | 18.18% | 0.00% | 0.00% | 9.09% | 18.18% | 0.00% | 0.00% |
| fear | CD | 27.27% | 36.36% | 0.00% | 0.00% | 27.27% | 36.36% | 27.27% | 0.00% |
| | CS | 9.09% | 18.18% | 0.00% | 0.00% | 9.09% | 18.18% | 9.09% | 0.00% |
| | UD | 9.09% | 18.18% | 0.00% | 0.00% | 9.09% | 18.18% | 9.09% | 0.00% |
| | US | 9.09% | 18.18% | 0.00% | 0.00% | 9.09% | 18.18% | 0.00% | 0.00% |
| disgust | CD | 18.18% | 36.36% | 0.00% | 0.00% | 36.36% | 27.27% | 27.27% | 9.09% |
| | CS | 9.09% | 27.27% | 0.00% | 0.00% | 9.09% | 18.18% | 9.09% | 0.00% |
| | UD | 9.09% | 18.18% | 0.00% | 0.00% | 9.09% | 18.18% | 18.18% | 0.00% |
| | US | 9.09% | 18.18% | 0.00% | 0.00% | 9.09% | 18.18% | 0.00% | 0.00% |
| anger | CD | 27.27% | 27.27% | 0.00% | 0.00% | 36.36% | 45.45% | 27.27% | 0.00% |
| | CS | 9.09% | 18.18% | 0.00% | 0.00% | 9.09% | 18.18% | 18.18% | 0.00% |
| | UD | 9.09% | 18.18% | 0.00% | 0.00% | 9.09% | 18.18% | 9.09% | 0.00% |
| | US | 9.09% | 18.18% | 0.00% | 0.00% | 9.09% | 18.18% | 0.00% | 0.00% |
| Task/Modality | | ASD | NT | ASD | NT | ASD | NT | ASD | NT |
| Legend 0% – 50% | | REC ET | | REC Button Clicks | | MIM ET | | MIM VT | |

*Note.* 'ET' and 'VT' denote eye tracking and video tracking, respectively. 'REC' denotes the recognition task, and 'MIM' denotes the mimicry task. Stimulus conditions 'US', 'UD', 'CS', and 'CD' denote uncustomized and static; uncustomized and dynamic; customized and static; and customized and dynamic, respectively.

condition, will add 15 DVs to the missing set, even if only 1 frame is missing out of the total possible 22. Also, our imputation study simulates higher levels of data loss (50% of samples missing for all DVs) than present in the full data set. The lowest MSE, RMSE, and MAE are achieved by KNN imputation. Therefore, we use KNN imputation to impute the missing values in the full data set.

## Group Discriminability

Considering a Type I error rate of α=0.05, we identify one candidate biomarker with Boruta: (MIM-US 'Disgust', ET %Gaze Face). To understand the partitioning of groups using (MIM-US 'Disgust', ET %Gaze Face), we fit the conditional inference tree (CIT) (Hothorn et al., 2015) shown in Figure 1. The CIT identifies a binary partitioning of samples based on (MIM-US 'Disgust', ET %Gaze Face) and tests the null hypothesis of independence between (MIM-US 'Disgust', ET %Gaze Face) and the group label. The null hypothesis is rejected with a p-value of 0.001. The CIT identifies the partition between groups as 0.53 with all participants in the ASD group, as well as three participants from the NT group, having a value of (MIM-US 'Disgust', ET %Gaze Face) that is greater than 0.53.

For r-Boruta, we determine the value of *perc* for our sample size to be 84.89%. Fifteen DVs are deemed important by r-Boruta, including (MIM-US 'Disgust', ET %Gaze Face) identified by Boruta. Table 5 reports these fifteen DVs along with their group means and standard deviations. Figure 2 shows the box plots by

Table 3

*Construct Validity for DVs by Task, Condition, and Expression*

| | Task/ Measure | REC %Gaze Face | REC %Acc | MIM %Gaze Face | MIM EXPR | AU 1 | AU 2 | AU 4 | AU 5 | AU 6 | AU 7 | AU 10 | AU 11 | AU 12 | AU 15 | AU 17 | AU 20 | AU 23 | AU 25 | AU 26 | AU 27 |
|---|---|---|---|---|---|---|---|---|---|---|---|---|---|---|---|---|---|---|---|---|---|
| surprise | CD | ✓ | ✓ | ✓ | ✓ | ✓ | ✓ | | ✓ | | | | | | | | | | ✓ | | ✓ |
| surprise | CS | ✓ | ✓ | ✓ | ✓ | ✓ | ✓ | | ✓ | | | | | | | | | | ✓ | | ✓ |
| surprise | UD | ✓ | ✓ | ✓ | ✓ | ✓ | ✓ | | ✓ | | | | | | | | | | ✓ | | ✓ |
| surprise | US | ✓ | ✓ | ✓ | ✓ | ✓ | ✓ | | ✓ | | | | | | | | | | ✓ | | ✓ |
| sad | CD | | ✓ | ✓ | ✓ | ✓ | | ✓ | | | | | | | ✓* | | | | | | |
| sad | CS | ✓ | ✓ | ✓ | | ✓ | | ✓ | | | | | | | ✓ | | | | | | |
| sad | UD | ✓ | ✓ | ✓ | ✓ | ✓ | | ✓ | | | | | | | ✓* | | | | | | |
| sad | US | ✓ | ✓ | ✓ | | ✓ | | ✓ | | | | | | | ✓ | | | | | | |
| happy | CD | ✓ | ✓ | ✓ | | | | | | ✓ | | | | ✓ | | | | | | | |
| happy | CS | ✓ | ✓ | ✓ | | | | | | ✓ | | | | ✓ | | | | | | | |
| happy | UD | ✓ | ✓ | ✓ | ✓ | | | | | | | | | ✓ | | | | | | | |
| happy | US | ✓ | ✓ | ✓ | | | | | | ✓ | | | | ✓ | | | | | | | |
| fear | CD | ✓ | ✓ | ✓ | | ✓ | ✓ | | ✓ | | | | | | | | | | ✓ | | |
| fear | CS | ✓ | ✓ | ✓ | | ✓ | ✓ | ✓ | ✓ | | | | | | | | ✓ | | ✓ | | |
| fear | UD | ✓ | ✓ | ✓ | | ✓ | ✓ | | ✓ | | | | | | | | | | ✓ | | |
| fear | US | ✓ | ✓ | ✓ | | ✓ | ✓ | | ✓ | | | | | | | | ✓ | | ✓ | | |
| disgust | CD | ✓ | ✓ | ✓ | ✓ | | | | | | | | | | | ✓ | | | | | |
| disgust | CS | ✓ | ✓ | ✓ | ✓ | | | | | | | | | | | ✓ | | | | | |
| disgust | UD | ✓ | ✓ | ✓ | ✓ | | | | | | | | | | | ✓ | | | | | |
| disgust | US | ✓ | ✓ | ✓ | ✓ | | | | | | | | | | | ✓ | | | | | |
| anger | CD | | ✓ | ✓ | ✓ | | | ✓ | | | ✓ | | | | | | | | | | |
| anger | CS | ✓ | ✓ | ✓ | ✓ | | | ✓ | | | ✓ | | | | | | | | ✓* | | |
| anger | UD | ✓ | ✓ | ✓ | ✓ | | | ✓ | | | ✓ | | | | | | | | | | |
| anger | US | ✓ | ✓ | ✓ | ✓ | | | ✓ | | | ✓ | | | | | | | | | | |

*Note.* *'✓' denotes a valid construct. For AUs, '✓' denotes a valid construct for the DVs associated with both AU ACT and ASYM, while '✓*' means that the DV associated with AU ACT is valid, but the DV associated with AU ASYM is not valid. A gray cell means that the AU is not part of the target expression. 'REC' denotes the recognition task, and 'MIM' denotes the mimicry task. Stimulus conditions 'US', 'UD', 'CS', and 'CD' denote uncustomized and static; uncustomized and dynamic; customized and static; and customized and dynamic, respectively.*

Table 4

*Comparison of Five Different Imputation Methods*

| Method | Mean MSE | Mean RMSE | Mean MAE |
|---|---|---|---|
| Mean Imputation | 0.1126 | 0.3144 | 0.2565 |
| Median Imputation | 0.1219 | 0.3259 | 0.2639 |
| MICE, Bayesian Ridge Regression | 0.2133 | 0.4228 | 0.3365 |
| MICE, Random Forest | 0.1139 | 0.3153 | 0.2580 |
| *KNN Imputation* | *0.0942* | *0.2845* | *0.2383* |

group for the fifteen DVs selected by Boruta and r-Boruta.

## Power Analysis

Figure 3 shows the PowerTools (Acharjee et al., 2020) output for estimated statistical power using a series of simulated samples of sizes 22, 44, 88, 176, 352, and 704 and the observed effect sizes for the fifteen DVs selected by Boruta and r-Boruta.

## Discussion

Our findings demonstrate the feasibility of DVs related to facial expression perception and production in

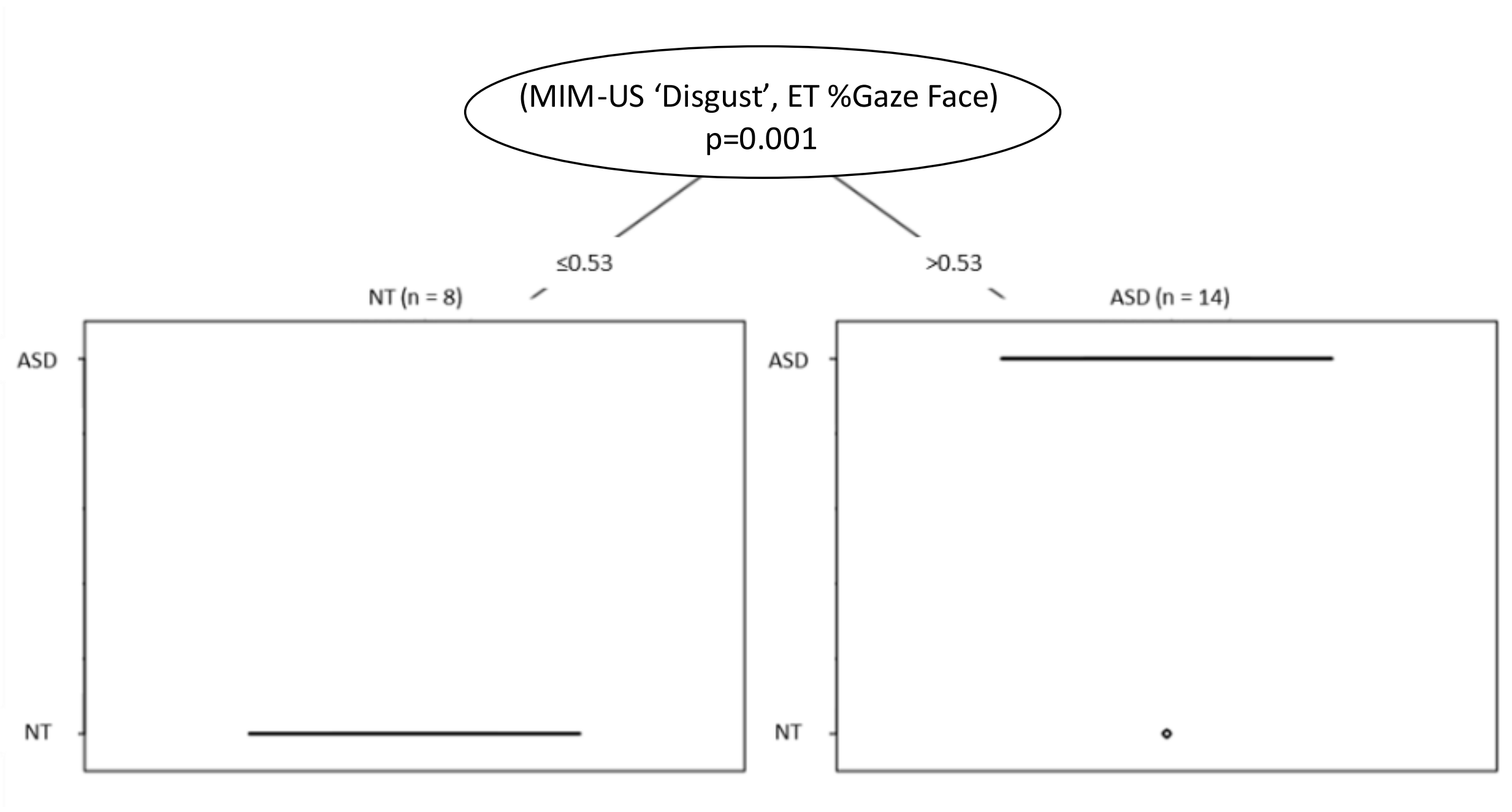

*Figure 1.* CIT of Group (ASD or NT) with Boruta-selected DV

Table 5

*Group Means and Standard Deviations for DVs selected by Boruta and r-Boruta*

| DV | ASD | NT |
|---|---|---|
| *(MIM-US 'Disgust', ET %Gaze Face)* | M = 0.77, SD = 0.16 | M = 0.49, SD = 0.13 |
| (MIM-UD 'Anger', VT ACT AU 7) | M = 0.12, SD = 0.17 | M = 0.34, SD = 0.33 |
| (MIM-CD 'Fear', VT ACT AU 5) | M = 0.37, SD = 0.29 | M = 0.22, SD = 0.34 |
| (MIM-CS 'Fear', VT ACT AU 5) | M = 0.35, SD = 0.34 | M = 0.14, SD = 0.17 |
| (MIM-US 'Fear', VT ACT AU 5) | M = 0.43, SD = 0.32 | M = 0.25, SD = 0.20 |
| (MIM-CD 'Happy', VT ACT AU 6) | M = 0.09, SD = 0.09 | M = 0.29, SD = 0.29 |
| (MIM-US 'Surprise', VT ACT AU 2) | M = 0.72, SD = 0.21 | M = 0.47, SD = 0.30 |
| (MIM-CS 'Fear', VT ASYM AU 25) | M = 0.37, SD = 0.33 | M = 0.12, SD = 0.18 |
| (MIM-CD 'Fear', VT ASYM AU 5) | M = 0.28, SD = 0.28 | M = 0.11, SD = 0.15 |
| (MIM-CS 'Fear', VT ASYM AU 5) | M = 0.30, SD = 0.29 | M = 0.13, SD = 0.19 |
| (MIM-US 'Surprise', VT ASYM AU 5) | M = 0.44, SD = 0.29 | M = 0.24, SD = 0.30 |
| (REC-CD 'Fear', ET %Gaze Face) | M = 0.50, SD = 0.18 | M = 0.34, SD = 0.08 |
| (REC-CS 'Fear', ET %Gaze Face) | M = 0.61, SD = 0.19 | M = 0.52, SD = 0.28 |
| (MIM-CS 'Sad', ET %Gaze Face) | M = 0.40, SD = 0.18 | M = 0.56, SD = 0.30 |
| (MIM-UD 'Sad', ET %Gaze Face) | M = 0.60, SD = 0.20 | M = 0.41, SD = 0.18 |

*Note.* 'ET' and 'VT' denote eye tracking and video tracking, respectively. 'REC' denotes the recognition task, and 'MIM' denotes the mimicry task. Stimulus conditions 'US', 'UD', 'CS', and 'CD' denote uncustomized and static; uncustomized and dynamic; customized and static; and customized and dynamic, respectively. M stands for the mean and SD for the standard deviation.

stratification biomarker discovery for ASD, considering the criteria of construct validity and group discriminability. With regard to construct validity (Table 3), invalid constructs may be due to limitations of the measurement tools (e.g., deep neural network models) or failure to elicit the expected response from NT participants (e.g., participants do not look at the face or do not produce the expression as expected).

For the tools, we use previously published stimuli and models for AU measurement (Witherow et al., 2024). Prior study of these tools report limitations for construct validity. In Witherow et al. (2024)'s original feasibility study of these tools, the following AUs are reported to have invalid constructs: AUs 5, 23, and 26 during 'anger'; AU 10 during 'disgust'; AU 20 during 'fear'; AU 6 during 'happy'; and AU 11 during 'sad'. Therefore, it is an expected finding that some of our constructs involving these same AUs (5, 6, 10, 11, 20, 23, and 26) are invalid. For the DVs involving EXPR measurements, some of the constructs for 'fear' (US, UD, CS, CD), 'happy' (US, CS, CD), and 'sad' (US, CS) are invalid. This may be due to differences in the image characteristics of our data set (e.g., lighting, backgrounds) or in the imitated expressions from our NT group compared to the prototypical expressions of the model's training data (Witherow et al., 2023).

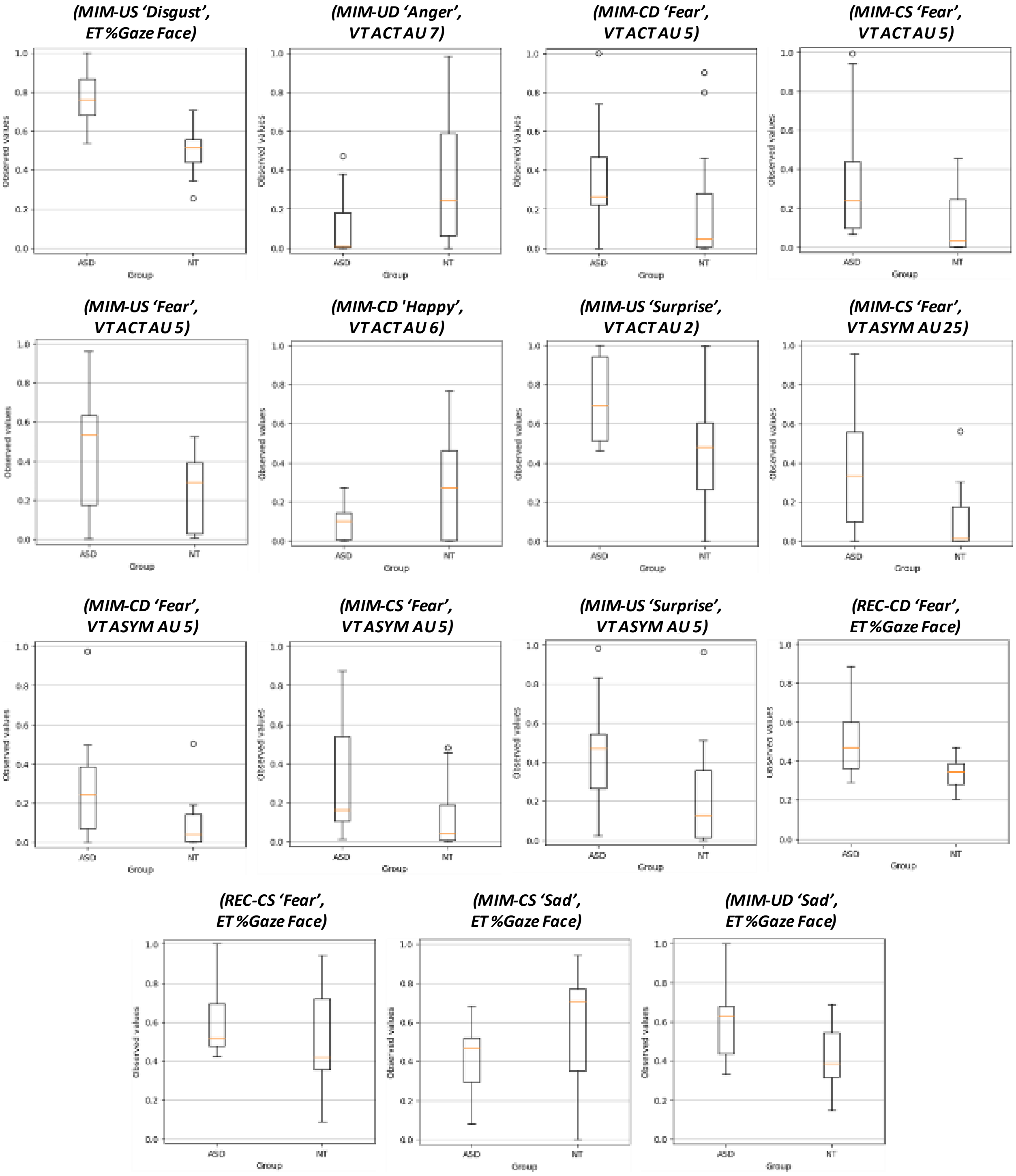

*Figure 2.* Box Plots for Fifteen DVs Selected by Boruta and r-Boruta

NT participant characteristics may also have an impact on construct validity. Studies of facial expression production with NT children have reported that negative expressions may be more difficult to elicit from children, including 'anger' (Grossard et al., 2018), 'fear' (Khan et al., 2019), and 'sad' (Grossard et al., 2018). Furthermore, Grossard et al. (2018) find that NT children produce significantly higher quality facial ex-pressions on request versus during imitation of an avatar and that the imitated expressions may be less credible and less recognizable. Therefore, some of the AUs and expressions in our study may be difficult to elicit especially among younger members of our age group. This may explain why we see invalid constructs for AU 25 'lips part' (ACT and ASYM for US, UD, CD; ASYM only for CS) during 'anger', and for AU 4 'brow lowerer' (ACT and ASYM for US, UD, CD) and AU 27 'mouth stretch' (ACT and ASYM for US, UD, CS, CD) during 'fear'. Also, the only two ET DV constructs that are not valid, (REC-CD 'Sad', ET %Gaze Face) and (REC-CD 'Anger', ET %Gaze Face), both occur at the end (CD condition) of the experiment and may be invalid due to waning attention from the some of the participants.

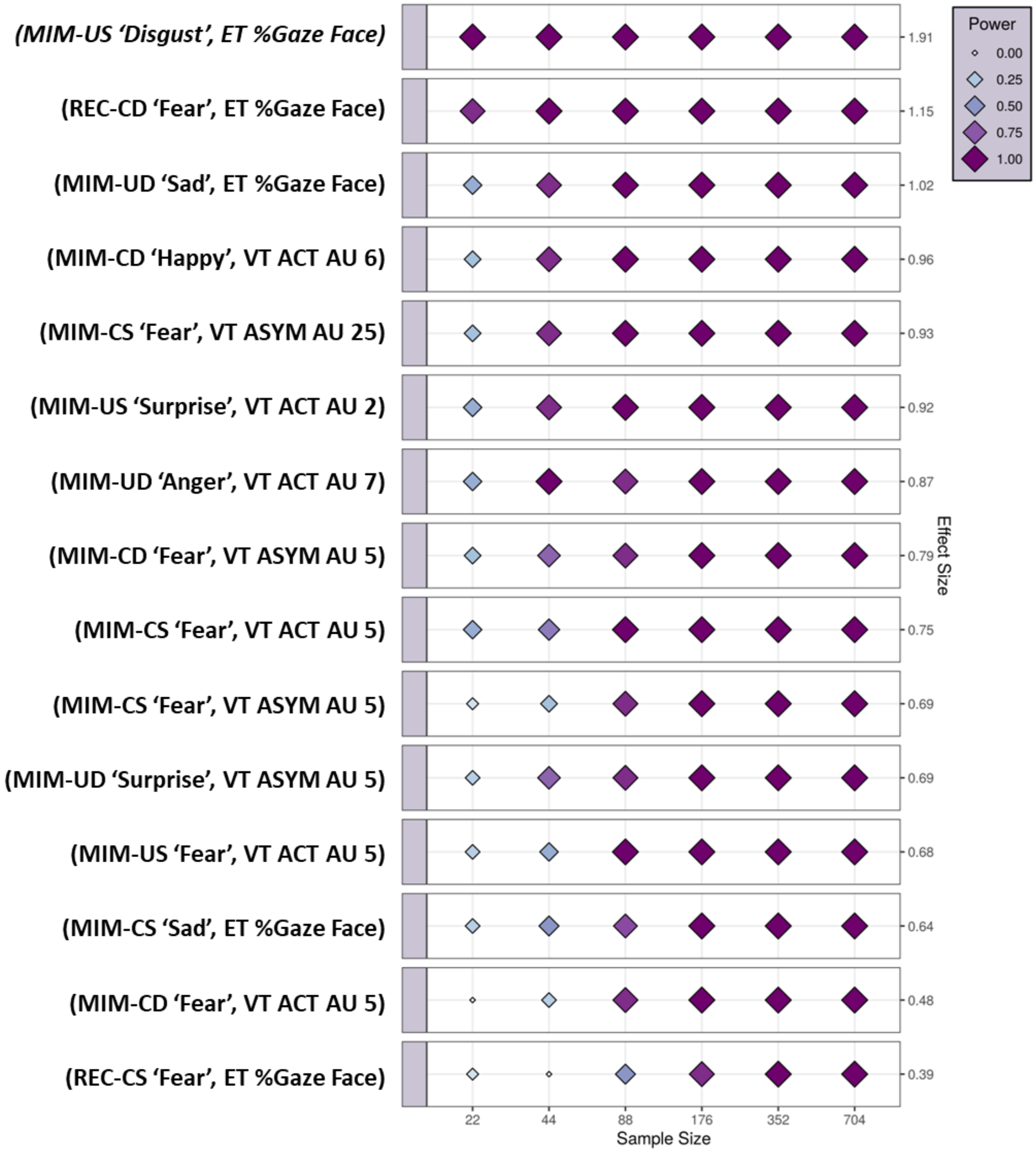

*Figure 3.* PowerTools Visualization of Observed Effect Sizes for Selected DVs and Estimated Statistical Power for Different Sample Sizes

Assessment of group discriminability criteria using Boruta has identified one candidate biomarker (MIM-US 'Disgust', ET %Gaze Face). (MIM-US 'Disgust', ET %Gaze Face) is greater for the ASD group, indicating that participants diagnosed with ASD spend more percentage gaze duration to the face while viewing the static 'disgust' expression presented on the uncustomized avatar compared to the NT group (Table 5). Our results are consistent with existing literature noting increased preference and engagement towards avatars among individuals diagnosed with ASD (Kellems et al., 2023; Putnam et al., 2019), as well as Pino et al. (2021)'s finding of higher gaze duration towards negative emotions. Using r-Boruta, we also identify fourteen additional DVs of interest, which we interpret with caution. No DVs involving %Acc or EXPR measurements are selected by either Boruta or r-Boruta. Like (MIM-US 'Disgust', ET %Gaze Face), the other ET DVs identified by r-Boruta ((REC-CD 'Fear', ET %Gaze Face), (REC-CS 'Fear', ET %Gaze Face), (MIM-CS 'Sad', ET %Gaze Face), and (MIM-UD 'Sad', ET %Gaze Face)) also involve negative emotions ('fear' or 'sad') and except for (MIM-CS 'Sad', ET %Gaze Face), report higher percentage gaze duration in the ASD group compared to the NT group (Table 5).

Six DVs involving ACT of AUs ((MIM-UD 'Anger', VT ACT AU 7), (MIM-CD 'Fear', VT ACT AU 5), (MIM-CS 'Fear', VT ACT AU 5), (MIM-US 'Fear', VT ACT AU 5), (MIM-CD 'Happy', VT ACT AU 6), (MIM-US 'Surprise', VT ACT AU 2)) are selected by r-Boruta. Four of these DVs ((MIM-CD 'Fear', VT ACT AU 5), (MIM-CS 'Fear', VT ACT AU 5), (MIM-US 'Fear', VT ACT AU 5), (MIM-US 'Surprise', VT ACT AU 2)) show greater activation, on average, by the ASD group compared to the NT group (Table 5). These findings corroborate prior mimicry studies that report more intense 'fear' (Faso et al., 2015) and 'surprise' (Volker et al., 2009) expressions by the ASD group relative to the NT group. In Manfredonia et al. (2019)'s study, group differences are also reported for 'fear' and 'surprise' expressions, specifically in AU 5 'upper lid raiser'. Manfredonia et al. (2019)'s ASD group shows lower activation of AU 5 on average relative to the NT group. However, Manfredonia et al. (2019) elicit facial responses using textual prompts rather than mimicry. On average, our ASD group has lower (MIM-UD 'Anger', VT ACT AU 7) and (MIM-CD 'Happy', VT ACT AU 6) compared to the NT group (Table 5). Prior studies have reported more intense (Faso et al., 2015) or no difference (Manfredonia et al., 2019; Volker et al., 2009) in the production of 'anger'. However, these prior studies use either static expressions (Faso et al., 2015; Volker et al., 2009) or a textual prompt (Manfredonia et al., 2019) to elicit facial responses. Recently, Keating et al. (2022) report significantly poorer recognition of dynamic 'anger' among their sample of participants diagnosed with ASD compared to an NT control group, even when controlling for alexithymia. It is possible that group differences in responses elicited by dynamic 'anger' stimuli may also extend to AU production and may be of interest for further study. Similar to Bangerter et al. (2020), we also find less activation of AU 6 during production of 'happy' ((MIM-CD 'Happy', VT ACT AU 6)), on average, in the ASD group compared to the NT group. We do not observe (Drimalla et al., 2021)'s finding of lower standard deviations of AU activations in the NT group compared to the ASD group. Heterogenous findings across our study and others may support the presence of subgroups of autistic participants that are more and less expressive (Quinde-Zlibut et al., 2022). However, these are preliminary findings and follow up studies are required.

For all five DVs involving ASYM of AUs ((MIM-CS 'Fear', VT ASYM AU 25), (MIM-CD 'Fear', VT ASYM AU 5), (MIM-CS 'Fear', VT ASYM AU 5), (MIM-UD 'Surprise', VT ASYM AU 5)) identified by r-Boruta, the ASD group shows higher asymmetry on average than the NT group (Table 5). Prior studies (Guha et al., 2015; Samad et al., 2015, 2016) on facial expression asymmetry in ASD have also found that individuals diagnosed with ASD may have higher levels of asymmetry in their expressions, specifically in the activation of left and right levator anguli oris muscles of the lower face (associated with AU 13 'cheek puffer', which is not present in the stimulus expressions) (Samad et al., 2015, 2016). In the lower face, we identify one DV that involves AU 25 'lips part', which may be associated with multiple facial muscles including depressor labii inferioris, mentalis, and orbicularis oris (Ekman et al., 2002). While prior studies focus on overall left-right asymmetry (Guha et al., 2015) or a few muscles of the lower face (Samad et al., 2015, 2016), we are able to capture muscle activations of the upper face as well. Three DVs related to asymmetry of AU 5 'upper lid raiser' activations identified in our analysis suggest that asymmetry of movements of the upper face may also be of interest for future study. Power analyses (Figure 3) show that maximum power is attained for (MIM-US 'Disgust', ET %Gaze Face) with our current sample size of 22.

## Limitations

Although our sample size is relatively small, the variance and covariance patterns in our sample have been managed by matching participants on age and gender. Further examination of larger data may shed additional information on perception and production behaviors. Our intuition in selecting robust DVs guided us in the use of statistical tools (Boruta and r-Boruta) that do not require independence or normality assumptions and work well for correlated data. Another direction could have been to use nonparametric or more general distribution functions (such as copula types). However, with the nonparametric approach, we may lose the dependence structure of the data, and with a copula-based approach, the dependence structure will be transformed based on the cumulative distribution function at the cost of interpretability. Further larger studies that incorporate deep phenotyping of participants and replication samples will move findings toward a more comprehensive understanding of these DVs and how they relate to IQ and other phenotypic variables (e.g., severity of ASD-related symptoms, social communication scores, adaptive function, etc.).

## Conclusion

This study demonstrates the feasibility of and provides a framework for ASD stratification biomarker discovery based on production and perception of facial expressions. We identify one candidate biomarker as well as fourteen additional DVs of interest and provide sample size recommendations for future studies. This study has found evidence of both more and less intense expressions in our ASD group, on average, depending on the stimulus type. More research is required to confirm and understand the significance of possible subgroups of autistic individuals based on these initial findings. Furthermore, we find several important DVs related to asymmetry of facial movements among individuals on the spectrum that we hope will facilitate follow up studies. Furthermore, we hope that this study will provide a foundation for larger studies involving deep phenotyping of participants and replication samples.

## References

Acharjee, A., Larkman, J., Xu, Y., Cardoso, V. R., & Gkoutos, G. V. (2020). A random forest based biomarker discovery and power analysis framework for diagnostics research. *BMC Medical Genomics*, *13*(1), 178. https://doi.org/10.1186/s12920-020-00826-6

Baltrusaitis, T., Zadeh, A., Lim, Y. C., & Morency, L. P. (2018). *OpenFace 2.0: Facial Behavior Analysis Toolkit* 2018 13th IEEE International Conference on Automatic Face & Gesture Recognition (FG 2018), https://ieeexplore.ieee.org/document/8373812

Bangerter, A., Chatterjee, M., Manfredonia, J., Manyakov, N. V., Ness, S., Boice, M. A., Skalkin, A., Goodwin, M. S., Dawson, G., Hendren, R., Leventhal, B., Shic, F., & Pandina, G. (2020). Automated recognition of spontaneous facial expression in individuals with autism spectrum disorder: parsing response variability. *Molecular Autism*, *11*(1), 31. https://doi.org/10.1186/s13229-020-00327-4

Bozgeyikli, L., Raij, A., Katkoori, S., & Alqasemi, R. (2018). A Survey on Virtual Reality for Individuals with Autism Spectrum Disorder: Design Considerations. *IEEE Transactions on Learning Technologies*, *11*(2), 133-151. https://doi.org/10.1109/TLT.2017.2739747

Cuve, H. C., Gao, Y., & Fuse, A. (2018). Is it avoidance or hypoarousal? A systematic review of emotion recognition, eye-tracking, and psychophysiological studies in young adults with autism spectrum conditions. *Research in Autism Spectrum Disorders*, *55*, 1-13. https://doi.org/https://doi.org/10.1016/j.rasd.2018.07.002

Drimalla, H., Baskow, I., Behnia, B., Roepke, S., & Dziobek, I. (2021). Imitation and recognition of facial emotions in autism: a computer vision approach. *Molecular Autism*, *12*(1), 27. https://doi.org/10.1186/s13229-021-00430-0

Egger, H. L., Dawson, G., Hashemi, J., Carpenter, K. L., Espinosa, S., Campbell, K., Brotkin, S., Schaich-Borg, J., Qiu, Q., & Tepper, M. (2018). Automatic emotion and attention analysis of young children at home: a ResearchKit autism feasibility study. *NPJ digital medicine*, *1*(1), 20.

Ekman, P., Friesen, W. V., & Hager, J. C. (2002). *Facial Action Coding System: Manual and Investigator's Guide*. Reseach Nexus.

Faso, D. J., Sasson, N. J., & Pinkham, A. E. (2015). Evaluating Posed and Evoked Facial Expressions of Emotion from Adults with Autism Spectrum Disorder. *Journal of Autism and Developmental Disorders*, *45*(1), 75-89. https://doi.org/10.1007/s10803-014-2194-7

Feliciano, P., Daniels, A. M., Snyder, L. G., Beaumont, A., Camba, A., Esler, A., Gulsrud, A. G., Mason, A., Gutierrez, A., & Nicholson, A. (2018). SPARK: A US cohort of 50,000 families to accelerate autism research. *Neuron*, *97*(3), 488-493.

Fombonne, E., Coppola, L., Mastel, S., & O'Roak, B. J. (2022). Validation of Autism Diagnosis and Clinical Data in the SPARK Cohort. *Journal of Autism and Developmental Disorders*, *52*(8), 3383-3398. https://doi.org/10.1007/s10803-021-05218-y

Ganin, Y., & Lempitsky, V. (2015). *Unsupervised domain adaptation by backpropagation* International conference on machine learning, Lille, France. https://dl.acm.org/doi/10.5555/3045118.3045244

Gibbs, V., Cai, R. Y., Aldridge, F., & Wong, M. (2021). Autism assessment via telehealth during the Covid 19 pandemic: Experiences and perspectives of autistic adults, parents/carers and clinicians. *Research in Autism Spectrum Disorders*, *88*, 101859. https://doi.org/https://doi.org/10.1016/j.rasd.2021.101859

Goerlich, K. S. (2018). The Multifaceted Nature of Alexithymia – A Neuroscientific Perspective [Perspective]. *Frontiers in Psychology*, *9*. https://doi.org/10.3389/fpsyg.2018.01614

Grossard, C., Chaby, L., Hun, S., Pellerin, H., Bourgeois, J., Dapogny, A., Ding, H., Serret, S., Foulon, P., Chetouani, M., Chen, L., Bailly, K., Grynszpan, O., & Cohen, D. (2018). Children Facial Expression Production: Influence of Age, Gender, Emotion Subtype, Elicitation Condition and Culture [Original Research]. *Frontiers in Psychology*, *9*. https://doi.org/10.3389/fpsyg.2018.00446

Guha, T., Yang, Z., Ramakrishna, A., Grossman, R. B., Darren, H., Lee, S., & Narayanan, S. S. (2015). On Quantifying Facial Expression-Related Atypicality of Children with Autism Spectrum Disorder. *Proc IEEE Int Conf Acoust Speech Signal Process*, *2015*, 803-807. https://doi.org/10.1109/icassp.2015.7178080

Hamidi, F., Gilani, N., Arabi Belaghi, R., Yaghoobi, H., Babaei, E., Sarbakhsh, P., & Malakouti, J. (2023). Identifying potential circulating miRNA biomarkers for the diagnosis and prediction of ovarian cancer using machine-learning approach: application of Boruta. *Frontiers in Digital Health*, *5*. https://doi.org/10.3389/fdgth.2023.1187578

Happé, F., & Frith, U. (2020). Annual Research Review: Looking back to look forward–changes in the concept of autism and implications for future research. *Journal of Child Psychology and Psychiatry*, *61*(3), 218-232.

Hothorn, T., Hornik, K., & Zeileis, A. (2015). ctree: Conditional inference trees. *The comprehensive R archive network*, *8*.

Kaneko, H. (2021). Examining variable selection methods for the predictive performance of regression models and the proportion of selected variables and selected random variables. *Heliyon*, *7*(6), e07356. https://doi.org/https://doi.org/10.1016/j.heliyon.2021.e07356

Kaufman, A. S., & Kaufman, N. L. (2004). *Kaufman brief intelligence test KBIT 2; manual* (2. ed.). Pearson Bloomington, Minn.

Keating, C. T., & Cook, J. L. (2020). Facial expression production and recognition in autism spectrum disorders: A shifting landscape. *Child and Adolescent Psychiatric Clinics*, *29*(3), 557-571.

Keating, C. T., Fraser, D. S., Sowden, S., & Cook, J. L. (2022). Differences Between Autistic and Non-Autistic Adults in the Recognition of Anger from Facial Motion Remain after Controlling for Alexithymia. *Journal of Autism and Developmental Disorders*, *52*(4), 1855-1871. https://doi.org/10.1007/s10803-021-05083-9

Kellems, R. O., Charlton, C. T., Black, B., Bussey, H., Ferguson, R., Gonçalves, B. F., Jensen, M., & Vallejo, S. (2023). Social Engagement of Elementary-Aged Children With Autism Live Animation Avatar Versus Human Interaction. *Journal of Special Education Technology*, *38*(3), 327-339. https://doi.org/10.1177/01626434221094792

Khan, R. A., Crenn, A., Meyer, A., & Bouakaz, S. (2019). A novel database of children's spontaneous facial expressions (LIRIS-CSE). *Image and Vision Computing*, *83-84*, 61-69. https://doi.org/https://doi.org/10.1016/j.imavis.2019.02.004

Kinnaird, E., Stewart, C., & Tchanturia, K. (2019). Investigating alexithymia in autism: A systematic review and meta-analysis. *Eur Psychiatry*, *55*, 80-89. https://doi.org/10.1016/j.eurpsy.2018.09.004

Kursa, M. B., Jankowski, A., & Rudnicki, W. R. (2010). Boruta – A System for Feature Selection. *Fundamenta Informaticae*, *101*, 271-285. https://doi.org/10.3233/FI-2010-288

Lee, H.-W., Chang, K., Uhm, J.-P., & Owiro, E. (2023). How Avatar Identification Affects Enjoyment in the Metaverse: The Roles of Avatar Customization and Social Engagement. *Cyberpsychology, Behavior, and Social Networking*, *26*(4), 255-262. https://doi.org/10.1089/cyber.2022.0257

Loth, E., Charman, T., Mason, L., Tillmann, J., Jones, E. J., Wooldridge, C., Ahmad, J., Auyeung, B., Brogna, C., & Ambrosino, S. (2017). The EU-AIMS Longitudinal European Autism Project (LEAP): design and methodologies to identify and validate stratification biomarkers for autism spectrum disorders. *Molecular Autism*, *8*(1), 1-19.

Loth, E., & Evans, D. W. (2019). Converting tests of fundamental social, cognitive, and affective processes into clinically useful biobehavioral markers for neurodevelopmental conditions. *Wiley Interdisciplinary Reviews: Cognitive Science*, *10*(5), e1499.

Loth, E., Garrido, L., Ahmad, J., Watson, E., Duff, A., & Duchaine, B. (2018). Facial expression recognition as a candidate marker for autism spectrum disorder: how frequent and severe are deficits? *Molecular Autism*, *9*(1), 1-11.

Manfredonia, J., Bangerter, A., Manyakov, N. V., Ness, S., Lewin, D., Skalkin, A., Boice, M., Goodwin, M. S., Dawson, G., Hendren, R., Leventhal, B., Shic, F., & Pandina, G. (2019). Automatic Recognition of Posed Facial Expression of Emotion in Individuals with Autism Spectrum Disorder. *Journal of Autism and Developmental Disorders*, *49*(1), 279-293. https://doi.org/10.1007/s10803-018-3757-9

McPartland, J. C., Bernier, R. A., Jeste, S. S., Dawson, G., Nelson, C. A., Chawarska, K., Earl, R., Faja, S., Johnson, S. P., & Sikich, L. (2020). The autism biomarkers consortium for clinical trials (ABC-CT): scientific context, study design, and progress toward biomarker qualification. *Frontiers in Integrative Neuroscience*, *14*, 16.

Meyer-Lindenberg, H., Moessnang, C., Oakley, B., Ahmad, J., Mason, L., Jones, E. J. H., Hayward, H. L., Cooke, J., Crawley, D., Holt, R., Tillmann, J., Charman, T., Baron-Cohen, S., Banaschewski, T., Beckmann, C., Tost, H., Meyer-Lindenberg, A., Buitelaar, J. K., Murphy, D. G., . . . Loth, E. (2022). Facial expression recognition is linked to clinical and neurofunctional differences in autism. *Molecular Autism*, *13*(1), 43. https://doi.org/10.1186/s13229-022-00520-7

Ness, S., Manyakov, N. V., Bangerter, A., Lewin, D., Jagannatha, S., Boice, M., Skalkin, A., Dawson, G., Goodwin, M. S., & Hendren, R. L. (2016). 1.32 THE JANSSEN AUTISM KNOWLEDGE ENGINE (JAKE™): A SET OF TOOLS AND TECHNOLOGIES TO ASSESS POTENTIAL BIOMARKERS FOR AUTISM SPECTRUM DISORDERS. *Journal of the American Academy of Child & Adolescent Psychiatry*, *10*(55), S110.

Papoutsaki, A., Sangkloy, P., Laskey, J., Daskalova, N., Huang, J., & Hays, J. (2016). *Webgazer: scalable webcam eye tracking using user interactions* Proceedings of the Twenty-Fifth International Joint Conference on Artificial Intelligence, New York, New York, USA.

Pellicano, E., Adams, D., Crane, L., Hollingue, C., Allen, C., Almendinger, K., Botha, M., Haar, T., Kapp, S. K., & Wheeley, E. (2024). Letter to the Editor: A possible threat to data integrity for online qualitative autism research. *Autism*, *28*(3), 786-792. https://doi.org/10.1177/13623613231174543

Pino, M. C., Vagnetti, R., Valenti, M., & Mazza, M. (2021). Comparing virtual vs real faces expressing emotions in children with autism: An eye-tracking study. *Education and Information Technologies*, *26*(5), 5717-32. https://doi.org/10.1007/s10639-021-10552-w

Putnam, C., Hanschke, C., Todd, J., Gemmell, J., & Kollia, M. (2019). Interactive Technologies Designed for Children with Autism: Reports of Use and Desires from Parents, Teachers, and Therapists. *ACM Trans. Access. Comput.*, *12*(3), 12. https://doi.org/10.1145/3342285

Quinde-Zlibut, J., Munshi, A., Biswas, G., & Cascio, C. J. (2022). Identifying and describing subtypes of spontaneous empathic facial expression production in autistic adults. *Neurodevelopmental Disorders*, *14*(1), 43. https://doi.org/10.1186/s11689-022-09451-z

Samad, M. D., Bobzien, J. L., Harrington, J. W., & Iftekharuddin, K. M. (2015). Analysis of facial muscle activation in children with autism using 3D imaging, 2015 IEEE International Conference on Bioinformatics and Biomedicine (BIBM), https://ieeexplore.ieee.org/document/7359704

Samad, M. D., Bobzien, J. L., Harrington, J. W., & Iftekharuddin, K. M. (2016). Non-intrusive optical imaging of face to probe physiological traits in autism spectrum disorder. *Optics & Laser Technology*, *77*, 221-228.

Shic, F., Barney, E. C., Naples, A. J., Dommer, K. J., Chang, S. A., Li, B., McAllister, T., Atyabi, A., Wang, Q., & Bernier, R. (2023). The Selective Social Attention task in children with autism spectrum disorder: Results from the Autism Biomarkers Consortium for Clinical Trials (ABC-CT) feasibility study. *Autism Research*, *16*(11), 2150-2159.

Shic, F., Naples, A. J., Barney, E. C., Chang, S. A., Li, B., McAllister, T., Kim, M., Dommer, K. J., Hasselmo, S., Atyabi, A., Wang, Q., Helleman, G., Levin, A. R., Seow, H., Bernier, R., Charwaska, K., Dawson, G., Dziura, J., Faja, S., . . . McPartland, J. C. (2022). The Autism Biomarkers Consortium for Clinical Trials: evaluation of a battery of candidate eye-tracking biomarkers for use in autism clinical trials. *Molecular Autism*, *13*(1), 15. https://doi.org/10.1186/s13229-021-00482-2

Simons Foundation. (2024). *About SPARK*. Simons Foundation,. Retrieved 03/24/2024 from https://sparkforautism.org/portal/page/about-spark/

Slade, E., & Naylor, M. G. (2020). A fair comparison of tree-based and parametric methods in multiple imputation by chained equations. *Statistics in Medicine*, *39*(8), 1156-1166. https://doi.org/https://doi.org/10.1002/sim.8468

Su, Q., Chen, F., Li, H., Yan, N., & Wang, L. (2018, 3-6 Dec. 2018). *Multimodal Emotion Perception in Children with Autism Spectrum Disorder by Eye Tracking Study* 2018 IEEE-EMBS Conference on Biomedical Engineering and Sciences (IECBES), https://ieeexplore.ieee.org/document/8626642

Toure, M., Klutse, N., Sarr, M., Kenne, A., Bhuiyan, M. A. E., Ndiaye, O., Daouda, B., Thiaw, W., Sy, I., Mbow, C., Sall, M., & Gaye, A. T. (2023). A New Multiple Imputation Approach Using Machine Learning to Enhance Climate Databases in Senegal. *ResearchGate*. https://doi.org/10.21203/rs.3.rs-3287168/v1

Troyanskaya, O., Cantor, M., Sherlock, G., Brown, P., Hastie, T., Tibshirani, R., Botstein, D., & Altman, R. B. (2001). Missing value estimation methods for DNA microarrays. *Bioinformatics*, *17*(6), 520-525. https://doi.org/10.1093/bioinformatics/17.6.520

Tsang, V. (2018). Eye-tracking study on facial emotion recognition tasks in individuals with high-functioning autism spectrum disorders. *Autism*, *22*(2), 161-170. https://doi.org/10.1177/1362361316667830

U.S. Food & Drug Administration. (2023, 03/22/2023). *Biomarker Qualification Submissions*. Retrieved 01/31/2024 from https://www.fda.gov/drugs/biomarker-qualification-program/biomarker-qualification-submissions

van Buuren, S., & Groothuis-Oudshoorn, K. (2011). mice: Multivariate Imputation by Chained Equations in R. *J. Statistical Software*, *45*(3), 1 - 67. https://doi.org/10.18637/jss.v045.i03

Volker, M. A., Lopata, C., Smith, D. A., & Thomeer, M. L. (2009). Facial encoding of children with high-functioning autism spectrum disorders. *Focus on Autism and Other Developmental Disabilities*, *24*(4), 195-204. https://doi.org/10.1177/1088357609347325

Vorst, H. C., & Bermond, B. (2001). Validity and reliability of the Bermond–Vorst alexithymia questionnaire. *Personality and individual differences*, *30*(3), 413-434.

Way, I. F., Applegate, B., Cai *, X., Franck, L. K., Black-Pond, C., Yelsma, P., Roberts, E., Hyter, Y., & Muliett, M. (2010). Children's Alexithymia Measure (CAM): A New Instrument for Screening Difficulties with Emotional Expression. *Journal of Child & Adolescent Trauma*, *3*(4), 303-318. https://doi.org/10.1080/19361521.2010.523778

Webb, S. J., Shic, F., Murias, M., Sugar, C. A., Naples, A. J., Barney, E., Borland, H., Hellemann, G., Johnson, S., Kim, M., Levin, A. R., Sabatos-DeVito, M., Santhosh, M., Senturk, D., Dziura, J., Bernier, R. A., Chawarska, K., Dawson, G., Faja, S., . . . Yuan, A. (2020). Biomarker Acquisition and Quality Control for Multi-Site Studies: The Autism Biomarkers Consortium for Clinical Trials [Methods]. *Frontiers in Integrative Neuroscience*, *13*. https://doi.org/10.3389/fnint.2019.00071

Witherow, M. A., Butler, C., Stedman, A., Shields, W., Ilgin, F., Diawara, N., Keener, J., Harrington, J. W., & Iftekharuddin, K. M. (2024). Customizable Avatars with Dynamic Facial Action Coded Expressions (CADyFACE) for Improved User Engagement. *ArXiv*. https://arxiv.org/abs/2403.07314

Witherow, M. A., Samad, M. D., Diawara, N., Bar, H. Y., & Iftekharuddin, K. M. (2023). Deep Adaptation of Adult-Child Facial Expressions by Fusing Landmark Features. *IEEE Transactions on Affective Computing*, 1-12. https://doi.org/10.1109/TAFFC.2023.3297075

Witherow, M. A., Shields, W. J., Samad, M. D., & Iftekharuddin, K. M. (2020). Learning latent expression labels of child facial expression images through data-limited domain adaptation and transfer learning. Applications of Machine Learning 2020, https://doi.org/10.1117/12.2569454